\documentstyle[12pt]{article}
\newcommand{\be}{\begin{eqnarray}}
\newcommand{\ee}{\end{eqnarray}}
\newcommand{\non}{\nonumber}
\newcommand{\dslash}{\hbox{$\partial$\kern-1.2ex\raise.2ex\hbox{$/$}}}
\newcommand{\Dslash}{\hbox{$D$\kern-1.5ex\raise.2ex\hbox{$/$}}}

\begin{document}

\begin{titlepage}
\strut\hfill UMTG--200
\vspace{.5in}
\begin{center}

\LARGE Fractional-Spin Integrals of Motion \\
for the Boundary Sine-Gordon Model \\
at the Free Fermion Point \\[1.0in]
\large Luca Mezincescu and Rafael I. Nepomechie\\[0.8in]
\large Physics Department, P.O. Box 248046, University of Miami\\[0.2in]  
\large Coral Gables, FL 33124 USA\\

\end{center}

\vspace{.5in}

\begin{abstract}
We construct integrals of motion (IM) for the sine-Gordon model with 
boundary at the free Fermion point ($\beta^2 = 4\pi$) which correctly 
determine the boundary $S$ matrix.  The algebra of these IM 
(``boundary quantum group'' at $q=1$) is a one-parameter family of 
infinite-dimensional subalgebras of twisted $\widehat{sl(2)}$.  We 
also propose the structure of the fractional-spin IM away 
from the free Fermion point ($\beta^2 \ne 4\pi$).
\end{abstract}
\end{titlepage}

\setcounter{footnote}{0}

\section{Introduction}

Much is known about trigonometric solutions of the Yang-Baxter 
equations, primarily because the algebraic structure underlying these 
equations has been identified and elucidated -- namely, quantum groups 
\cite{qgroups}.  In contrast, much less is known about corresponding       
solutions of the {\it boundary} Yang-Baxter equation \cite{cherednik}, 
because the relevant algebraic structure (some sort of ``boundary 
quantum group'') has not yet been formulated.

Motivated in part by such considerations, we have undertaken 
together with A.  B.  Zamolodchikov 
a project \cite{MNZ} to construct fractional-spin integrals of 
motion \cite{luscher} - \cite{BL} of the sine-Gordon model with 
boundary \cite{GZ}, which should generate precisely such an algebraic 
structure.  A further motivation for this work is to determine the 
exact relation between the parameters of the action and the parameters 
of the boundary $S$ matrix given in \cite{GZ}.

We focus here on the special case of the free Fermion point ($\beta^2 
= 4\pi$) \cite{coleman}.  We construct Fermionic integrals of motion 
(IM) which correctly determine the boundary $S$ matrix, and which flow 
to the ``topological charge'' as the ``boundary magnetic field'' $h$ 
is varied from $h=0$ to $h \rightarrow \infty$.  The algebra of these 
IM (``boundary quantum group'' at $q=1$) is a one-parameter ($h$) 
family of infinite-dimensional subalgebras of twisted 
$\widehat{sl(2)}$.  We also give an Ansatz for the structure of the 
fractional-spin IM away from the free Fermion point ($\beta^2 \ne 4\pi$).  
A preliminary account of this work has appeared in Refs.  \cite{MN1} and 
\cite{MN2}.

We now outline the contents of the paper.  In Sec.  2 we briefly 
review some basic properties of both bulk and boundary sine-Gordon 
field theory for general values of $\beta$.  In Sec.  3 we consider 
the special case of the free Fermion point.  We first review the 
Fermionic description of the fractional-spin IM for the bulk theory.  
There follows the main part of the paper, in which we generalize these 
results to the boundary theory.  In particular, we describe the 
Fermionic IM and their algebra, and we verify that these IM correctly 
determine the boundary $S$ matrix.  In Sec.  4 we  propose the structure 
of the fractional-spin IM away from the free Fermion point. We conclude 
in Sec. 5 with a brief discussion of some open problems.
 
\section{Sine-Gordon field theory}

We briefly review here some basic properties of the sine-Gordon field 
theory for general values of $\beta$.  We first consider the bulk 
theory.  We describe the soliton/antisoliton $S$ matrix \cite{ZZ} and the 
fractional-spin integrals of motion \cite{BL}. We then turn to the boundary 
sine-Gordon theory, and describe the boundary $S$ matrix \cite{GZ}.

\subsection{Bulk theory}

The Lagrangian density of the bulk sine-Gordon field theory in 
Minkowski spacetime is given by
\be
{\cal L}_0 = -{1\over 2} (\partial_\mu \Phi)^2 + {m_0^2\over \beta^2} 
\cos ( \beta \Phi) \,,
\label{bulkSG}
\ee
where $\Phi(x,t)$ is a real scalar field, $m_0$ has dimensions of mass,  
and $\beta$ is a dimensionless coupling constant. For $4\pi \le 
\beta^2 \le 8\pi$, the particle spectrum consists only of solitons and 
antisolitons, with equal masses $m$, and with ``topological charge''
\be
T = {\beta\over 2\pi}\int_{-\infty}^\infty dx {\partial\over \partial 
x} \Phi(x,t)
\label{sgtopcharge}
\ee
equal to $+1$ or $-1$, respectively. The particles' two-momenta 
$p_\mu$ are conveniently parameterized in terms of their rapidities 
$\theta$: 
\be
p_0 = m \cosh \theta  \,, \qquad p_1 = m \sinh \theta \,.
\label{rapidity}
\ee

This theory is integrable; i.e., it possesses an infinite number of 
integer-spin integrals of motion.  Consequently, the scattering of 
solitons is factorizable.  (See \cite{ZZ} and references therein.) For 
the case of two particles with rapidities $\theta_1$ and $\theta_2$, 
the two-particle $S$ matrix $S(\theta)$ (where $\theta = \theta_1 - 
\theta_2$) is defined by
\be
A_{a_1}(\theta_1)^\dagger \ A_{a_2}(\theta_2)^\dagger =
S_{a_1\ a_2}^{b_1\ b_2}(\theta) \ A_{b_2}(\theta_2)^\dagger \
A_{b_1}(\theta_1)^\dagger \,,
\ee
where $A_\pm(\theta)^\dagger$ are ``particle-creation operators''.
The $S$ matrix obeys the Yang-Baxter equation
\be
S_{12}(\theta)\ S_{13}(\theta + \theta')\ S_{23}(\theta')
= S_{23}(\theta')\ S_{13}(\theta + \theta')\ S_{12}(\theta) \,,
\ee 
and therefore has the form
\be
S(\theta) = \rho (\theta)
\left(\begin{array}{cccc}
     a(\theta) & 0         & 0         & 0 \\
      0        & b(\theta) & c(\theta) & 0 \\
      0        & c(\theta) & b(\theta) & 0 \\ 
      0        & 0         & 0         & a(\theta) \\
      \end{array} \right)   \,,
\label{ZZ}
\ee
where
\be
a(\theta) &=& \sin[ \lambda (\pi - u)] \non \\
b(\theta) &=& \sin (\lambda u) \non \\
c(\theta) &=& \sin (\lambda \pi) 
\ee
and $u=-i \theta$. The relation between the coupling constant $\beta$ 
and the parameter $\lambda$ of the $S$ matrix
\be
\lambda = {8\pi\over \beta^2} - 1 
\label{relofparams}
\ee
can be inferred \cite{ZZ} from semiclassical results for the 
\mbox{$S$ matrix} and the mass spectrum.  The unitarization factor 
$\rho(\theta)$ can be found in Ref \cite{ZZ}.

In addition to having an infinite number of integrals of motion (IM) 
of integer spin, the sine-Gordon model also has fractional-spin IM.  
Within the framework of Euclidean-space perturbed conformal field 
theory \cite{Z2}, the IM $Q_\pm$ and $\bar Q_\pm$ (with Lorentz spin 
$\lambda$ and
$-\lambda$, respectively) are given by \cite{BL}
\be
Q_\pm & = & {1\over 2 \pi i}\left( \int dz J_{\pm} 
+ \int d\bar z H_{\pm} \right) \,, \non \\
\bar Q_\pm & = & {1\over 2 \pi i}\left( \int d\bar z \bar J_{\pm} 
+ \int dz \bar H_{\pm} \right) \,,
\label{sgIM}
\ee
where, up to a normalization factor, 
\be
J_{\pm}(x\,, t) & = & \exp \left[
\pm {2 i\over \hat \beta} \phi(x\,, t) \right] \,, \qquad 
\bar J_{\pm}(x\,, t)  = \exp \left[
\mp {2 i\over \hat \beta} \bar \phi(x\,, t) \right] \,,
\non \\
H_{\pm}(x\,, t) & = & m_{0}{\hat\beta^{2}\over \hat\beta^{2} - 2} 
\exp \left[
\pm i \left( {2\over \hat\beta} - \hat\beta \right) \phi(x\,, t)
\mp i \hat\beta \bar \phi(x\,, t)\right] \,, \non \\
\bar H_{\pm}(x\,, t) & = &  m_{0}{\hat\beta^{2}\over \hat\beta^{2} - 2} 
\exp \left[
\mp i \left( {2\over \hat\beta} - \hat\beta \right) \bar \phi(x\,, t)
\pm i \hat\beta  \phi(x\,, t) \right] \,,
\ee 
with $\hat\beta = \beta/ \sqrt{4\pi}$ and where $\phi(x,t)$ and 
$\bar \phi(x,t)$ are given by
\be
\phi(x,t) &= &{\sqrt{4\pi}\over 2}\left( \Phi(x,t) + \int_{-\infty}^x dy \ 
\partial_t  \Phi(y,t) \right) \,, \non \\
\bar \phi(x,t) &= &{\sqrt{4\pi}\over 2}\left( \Phi(x,t) - \int_{-\infty}^x dy \ 
\partial_t  \Phi(y,t) \right) \,, 
\ee
respectively. An appropriate regularization prescription is implicit 
in the above expressions for the currents.

These IM obey the so-called $q$-deformed twisted affine $sl(2)$ 
algebra ($\widehat{sl_q(2)}$) with zero center, where 
\be 
q= e^{-i \pi (\lambda - 1)} \,. 
\ee
Moreover, these IM have nontrivial commutation relations with the 
particle-creation operators, which in terms of the matrix notation
$A(\theta)^\dagger = \left( \begin{array}{l}
             A_+(\theta)^\dagger \\
             A_-(\theta)^\dagger 
             \end{array} \right)$ 
are of the form 
\be
Q_\pm A(\theta)^\dagger &=&  q^{\pm \sigma_3} A(\theta)^\dagger 
Q_\pm + \alpha  e^{\lambda \theta} \sigma_\mp A(\theta)^\dagger 
\,,  \non \\         
\bar Q_\pm A(\theta)^\dagger &=&  q^{\mp \sigma_3} A(\theta)^\dagger 
\bar Q_\pm + \bar\alpha  e^{-\lambda \theta} \sigma_\mp  
A(\theta)^\dagger \,,  \non \\
T A(\theta)^\dagger &=& A(\theta)^\dagger T + \sigma_3 A(\theta)^\dagger 
\,, \label{comrel}
\ee
where the values of $\alpha$ and $\bar\alpha$ depend on the 
normalization and regularization of the currents, with
\be
\alpha \rightarrow 1 \,, \quad \bar\alpha \rightarrow 1 \quad
{\hbox{ for }} \quad q \rightarrow 1 \,.
\label{alphabeta}
\ee
Associativity of the products in $Q A_{a_1}(\theta_1)^\dagger 
A_{a_2}(\theta_2)^\dagger |0\rangle$ and invariance of the vacuum 
$Q |0\rangle =0$ (where $Q = Q_\pm \,, T$ or $Q = \bar Q_\pm  \,, T$ ); 
or, equivalently,
\be
\left[ \check S \,, \Delta (Q) \right] = 0
\ee
(where $\check S$ is the ``$S$ operator'' and $\Delta$ is the 
comultiplication) leads to the $S$ matrix (\ref{ZZ}), up to the 
unitarization factor. Note that this derivation of the $S$ 
matrix yields the relation (\ref{relofparams}) between the coupling 
constant $\beta$ and the parameter $\lambda$ of the $S$ matrix.
Also note that only half of the fractional-spin IM are needed to 
determine the $S$ matrix.
This approach of computing $S$ matrices is a realization in quantum field theory
of the general program \cite{qgroups} of linearizing the problem of finding
trigonometric solutions of the Yang-Baxter equations.

\subsection{Boundary theory}

Following Ghoshal and Zamolodchikov \cite{GZ}, we consider the sine-Gordon 
field theory on the negative half-line $x \le 0$, with action
\be
S = \int_{-\infty}^\infty dt \left\{ \int_{-\infty}^0 dx\ {\cal L}_0 + 
M \cos {\beta\over 2}(\Phi - \Phi_0) \Big\vert_{x=0} \right\} \,,
\label{boundarySG}
\ee
where ${\cal L}_0$ is given by Eq.  (\ref{bulkSG}), and $M$ and 
$\Phi_0$ are parameters which specify the boundary conditions.  There 
is compelling evidence that this choice of boundary conditions 
preserves the integrability of the bulk theory, and hence, the 
scattering of the solitons remains factorizable.  The boundary $S$ 
matrix $R(\theta)$, which is defined by \cite{GZ}
\be 
A(\theta)^\dagger B = R(\theta) A(-\theta)^\dagger B \,,
\label{boundarySmatrix-definition}
\ee 
where $B$ is the boundary operator, describes the scattering of 
solitons and antisolitons off the boundary.  The boundary $S$ matrix 
must satisfy the boundary Yang-Baxter equation \cite{cherednik}
\be
S_{12}(\theta_{1}-\theta_{2})\ R_{1}(\theta_{1})\ S_{21}(\theta_{1}+\theta_{2})\
R_{2}(\theta_{2}) =
R_{2}(\theta_{2})\ S_{12}(\theta_{1}+\theta_{2})\ R_{1}(\theta_{1})\ 
S_{21}(\theta_{1}-\theta_{2}) \,,
\ee 
and therefore has the form \cite{GZ}, \cite{devega}
\be
R(\theta) =	r( \theta)
\left(\begin{array}{cc}
\cosh ( \lambda\theta + i\xi)	& -{i\over 2} k_+ \sinh( 2\lambda\theta)	\\
		-{i\over 2} k_- \sinh( 2\lambda\theta) &  \cosh ( \lambda\theta - i\xi) 
	  \end{array} \right)	\,.
\label{boundarySmatrix}
\ee
Evidently, this matrix depends on the three boundary parameters $\xi$, $k_+$, 
$k_-$; however, by a suitable gauge transformation, the off-diagonal
elements can be made to coincide. Thus, the boundary $S$ matrix in 
fact depends on only two boundary parameters, as does the action 
(\ref{boundarySG}). The relation between these two sets of parameters
is not known for general values of $\beta$.

\section{Free Fermion point}

As discussed in the Introduction, we would like to construct 
fractional-spin IM for the sine-Gordon field theory with boundary for 
general values of $\beta$.  We consider in this Section the special 
case $\beta^2 = 4\pi$ which corresponds to $q=1$ (i.e., undeformed 
algebra).  We begin by reviewing the Fermionic description of the 
fractional-spin IM for the bulk theory \cite{leclair}. We then generalize 
these results to the boundary theory.  In particular, we describe the 
Fermionic IM and their algebra, and we verify that these IM correctly 
determine the boundary $S$ matrix. 

\subsection{Bulk theory}

As is well known \cite{coleman}, the bulk sine-Gordon field 
theory (\ref{bulkSG}) with $\beta^2 = 4\pi$ is equivalent to a free massive 
Dirac field theory. The corresponding Lagrangian density is\footnote{Our 
conventions are $\eta^{00}= - 1 = - \eta^{11}$; 
$\bar\Psi = \Psi^{\dagger T} \gamma^0$;
$\gamma^0 = -i \sigma_2$; $\gamma^1 = \sigma_1$, where $\sigma_i$ are
the Pauli matrices; $\partial_{\pm} = {1\over 2}(\pm \partial_{0} + 
\partial_{1})$}
\be
{\cal L}_0^{FF} &=& {i\over 2} \bar \Psi 
\buildrel \leftrightarrow \over \dslash \Psi
- i m \bar \Psi \Psi   \non \\
           &=& -i \left[ \bar\psi_- \partial_+ \bar\psi_+ 
+ \bar\psi_+ \partial_+ \bar\psi_- - \psi_- \partial_- \psi_+ 
- \psi_+ \partial_- \psi_- - m ( \bar\psi_- \psi_+ - \psi_-\bar\psi_+ )
\right] \,, 
\label{bulk}
\ee
where $\Psi = \left( \begin{array}{l}
                       \bar\psi_+ \\
                           \psi_+ 
               \end{array} \right)$,
$\Psi^\dagger = \left( \begin{array}{l}
                      \bar\psi_- \\
                          \psi_- 
               \end{array} \right)$.
Evidently, the Lagrangian has a $U(1)$ symmetry; $\psi_+$, 
$\bar\psi_+$, have charge $+1$ and $\psi_-$, $\bar\psi_-$ have 
charge $-1$. 

By solving the field equations
\be
\partial_{-} \psi_{\pm} = {m\over 2} \bar \psi_{\pm} \,, \qquad \qquad
\partial_{+} \bar \psi_{\pm} = {m\over 2} \psi_{\pm} \,,
\label{fieldequations}
\ee
we are led (following the conventions of \cite{GZ}) to the mode 
expansions
\be
\psi_+(x,t) &=& \sqrt{m\over 4\pi} 
\int_{-\infty}^\infty d\theta\ e^{{\theta\over 2}}
    \left[ \omega A_-(\theta) e^{i p \cdot x}
      + \omega^* A_+(\theta)^\dagger e^{-i p \cdot x} \right] \,, \non \\
\bar\psi_+(x,t) &=& \sqrt{m\over 4\pi} 
\int_{-\infty}^\infty d\theta\ e^{-{\theta\over 2}}
    \left[\omega^* A_-(\theta) e^{i p \cdot x}
        + \omega A_+(\theta)^\dagger e^{-i p \cdot x} \right] \,,
\label{modes}
\ee
where $p \cdot x = p_0 t + p_1 x$ with $p_{\mu}$ given by Eq. 
(\ref{rapidity}), and $\omega = e^{i \pi/4} \,, \omega^* = e^{-i \pi/4}$.
Canonical quantization implies that the only nonvanishing anticommutation 
relations for the modes are
\be 
\left\{ A_\pm (\theta) \,, A_\pm (\theta')^\dagger \right\}
= \delta (\theta - \theta') \,.
\ee
We regard $A_\pm (\theta)^\dagger$ as creation operators.

The fractional-spin integrals of motion (\ref{sgIM}) and the 
topological charge (\ref{sgtopcharge}) have at the free Fermion point 
the form \cite{leclair}
\be
Q_\pm & = & -{i\over m} \int_{-\infty}^{\infty} dx\ \psi_\pm \dot \psi_\pm 
= \int_{-\infty}^{\infty} d\theta\ e^{\theta} 
      A_{\pm}(\theta)^{\dagger} A_{\mp}(\theta) \,, \non \\
\bar Q_\pm  & = & -{i\over m} \int_{-\infty}^{\infty} 
dx\ \bar \psi_\pm \dot {\bar \psi}_\pm 
= \int_{-\infty}^{\infty} d\theta\ e^{-\theta} 
      A_{\pm}(\theta)^{\dagger} A_{\mp}(\theta) \,, \non \\
T & = & - \int_{-\infty}^{\infty} dx\ 
\left( \psi_-\psi_+ + \bar\psi_-\bar\psi_+ \right) 
= \int_{-\infty}^{\infty} d\theta\ \left[
A_{+}(\theta)^{\dagger} A_{+}(\theta) - A_{-}(\theta)^{\dagger} A_{-}(\theta)
\right] \,,
\label{fermicharges}
\ee
where the dot $(\ \dot{}\ )$ denotes differentiation with respect to time. 
Here we normalize the IM so that their commutation relations with 
the particle-creation operators are given by Eq. (\ref{comrel}) with 
$\alpha = \bar \alpha = 1$.

Let us set $Q^\pm_{-1} \equiv Q_\pm$, $Q^\pm_{1} \equiv \bar Q_\pm$, 
$T_0 \equiv T$. By closing the algebra of these IM, we obtain the 
infinite-dimensional twisted affine Lie algebra $\widehat{sl(2)}$ with 
zero center \cite{leclair}:
\be
\left[ Q_n^+ \,, Q_m^- \right] & = &T_{n+m} \,, \non \\
\left[ T_n \,, Q_m^\pm \right] & = & \pm 2 Q_{n+m}^\pm \,,   
\label{bulkalgebra}
\ee
(all other commutators vanish), where
\begin{equation}
\left.
\begin{array}{rcl}
Q^+_n & = & \int_{-\infty}^\infty d\theta\ e^{-n\theta} 
A_+(\theta)^\dagger A_-(\theta)  \\ 
Q^-_n & = & \int_{-\infty}^\infty d\theta\ e^{-n\theta} 
A_-(\theta)^\dagger A_+(\theta) 
\end{array}
\right\}\,, \qquad n {\hbox{ odd}} \,,
\label{bulkbasis1}   
\end{equation}
\be
T_n & = & \int_{-\infty}^\infty d\theta\ e^{-n\theta} 
\left[ A_+(\theta)^\dagger A_+(\theta) 
-  A_-(\theta)^\dagger A_-(\theta) \right] \,, 
\qquad n {\hbox{ even}} \,. 
\label{bulkbasis2} 
\ee
Note that
\be
{Q_n^+}^\dagger = Q_n^- \,, \qquad \qquad T_n^\dagger= T_n \,.
\ee  
The expressions in coordinate space for $Q^{\pm}_{n}$ and $T_n$ have 
derivatives of highest order $|n|$.

\subsection{Boundary theory}

The sine-Gordon field theory with boundary (\ref{boundarySG})
for $\beta^2 = 4\pi$ is equivalent to \cite{MN1}, \cite{AKL}
\be
S = \int_{-\infty}^\infty dt \left\{ \int_{-\infty}^0 dx\ {\cal L}_0^{FF}
+ L_{boundary} \right\} \,,
\label{action}
\ee
where ${\cal L}_0^{FF}$ is given by Eq. (\ref{bulk}), 
and $L_{boundary}$ is given by 
\be
L_{boundary} &=& {i\over 2} \Big[ 
e^{i(\phi + \varphi)} \bar\psi_+ \psi_+
+ e^{-i(\phi + \varphi)} \bar\psi_- \psi_- + a \dot a \non \\
& & \mbox{} - h \left( e^{i \varphi} \bar\psi_+ + e^{-i \varphi} \bar\psi_-
+ e^{i \phi} \psi_+ + e^{-i \phi} \psi_- \right) a \Big]\Big\vert_{x=0} \,,
\label{boundary}
\ee
where $a(t)$ is a Fermionic boundary degree of freedom, and $h$, 
$\phi$, $\varphi$ are real parameters which specify the boundary 
conditions.  Varying the action gives (upon eliminating $a(t)$ through 
its equations of motion) the following boundary conditions
\be
\left( e^{i \varphi} \bar\psi_+ - e^{-i \varphi} \bar\psi_-
+ e^{i \phi} \psi_+ - e^{-i \phi} \psi_- \right)\Big\vert_{x=0} = 0 \,,
\label{BC-interpolating1}
\ee
\be
& & {d\over dt}\left( e^{i \varphi} \bar\psi_+ + e^{-i \varphi} \bar\psi_-
- e^{i \phi} \psi_+ - e^{-i \phi} \psi_- \right)\Big\vert_{x=0} \non \\
& & \quad \mbox{} - h^2  \left( e^{i \varphi} \bar\psi_+ 
+ e^{-i \varphi} \bar\psi_-
+ e^{i \phi} \psi_+ + e^{-i \phi} \psi_- \right)\Big\vert_{x=0} = 0 \,.
\label{BC-interpolating2}
\ee

This action is a generalization of the one for the off-critical Ising 
field theory (free massive Majorana field) with a boundary magnetic 
field $h$ \cite{GZ}.  The action given in \cite{AKL} differs slightly from 
(\ref{boundary}); specifically, it involves two Fermionic boundary 
degrees of freedom instead of one.  However, it gives the same 
boundary conditions (\ref{BC-interpolating1}), 
(\ref{BC-interpolating2}), apart from minor differences in notation.  
The relation between the parameters of the Bosonic and Fermionic 
actions is discussed in \cite{AKL}.

Computing the boundary $S$ matrix $R(\theta)$ according to the 
definition (\ref{boundarySmatrix-definition}) and the mode expansions 
(\ref{modes}) along the lines of Ref.  \cite{GZ}, we obtain the result 
(\ref{boundarySmatrix}), with $\lambda = 1$ and
\be
e^{2i \xi} &=& {e^{-i(\phi - \varphi)} - {m\over h^2}\over
e^{i(\phi - \varphi)} - {m\over h^2}} \,, 
\label{iden1} \\
k_\pm &=&- {m e^{\mp i(\phi + \varphi)}\over h^2 
{\sqrt{1 - {2m\over h^2}\cos(\phi - \varphi) + {m^2\over h^4}}}}\,,  
\label{iden2} \\
r(\theta) &=& {{\sqrt{1 - {2m\over h^2}\cos(\phi - \varphi) 
+ {m^2\over h^4}}}
\over \cosh(\phi - \varphi) - i\sinh\theta -{m\over h^2}\cosh^2 
\theta} \,.
\ee
Indeed, the action (\ref{action}), (\ref{boundary}) was formulated 
specifically to reproduce this result of \cite{GZ}.

We turn now to the problem of constructing IM.
In analogy with the bulk IM (\ref{fermicharges}), we define
the following charges on the half-line:
\be
Q_\pm &=& -{i\over m} \int_{-\infty}^0 dx\ \psi_\pm \dot \psi_\pm  
\,, \non \\   
\bar Q_\pm  &=& -{i\over m}  \int_{-\infty}^0 dx\ 
\bar \psi_\pm \dot {\bar \psi}_\pm \,, \non \\ 
T &=& - \int_{-\infty}^0 dx\ \left( \psi_-\psi_+ + \bar\psi_-\bar\psi_+ 
\right) \,.
\label{fermichargesbound}
\ee
Evidently, these charges are of the form 
$\int_{-\infty}^0 dx\ J^0$, where $J^0$ is the time component of a 
two-component current $J^\mu$ which is conserved ($\partial_\mu 
J^\mu = 0$). Nevertheless, in general these charges (unlike 
$\int_{-\infty}^\infty dx\ J^0$) are {\it not} separately conserved, since
$J^1 \big\vert_{x=0} \ne 0$. 

For future convenience, we also define (in analogy with (\ref{bulkbasis1}), 
(\ref{bulkbasis2})) the quantities
\begin{equation}
\left.
\begin{array}{rcl}
Q^+_n & = & \int_0^{\infty} d\theta\ e^{-n\theta} 
A_+(\theta)^\dagger A_-(\theta)  \\ 
Q^-_n & = & \int_0^{\infty} d\theta\ e^{-n\theta} 
A_-(\theta)^\dagger A_+(\theta) 
\end{array}
\right\}\,, \qquad n {\hbox{ odd}} \,,
\label{boundbasis1}   
\end{equation}
\be
T_n & = & \int_0^{\infty} d\theta\ e^{-n\theta} 
\left[ A_+(\theta)^\dagger A_+(\theta) 
-  A_-(\theta)^\dagger A_-(\theta) \right] \,, 
\qquad n {\hbox{ even}} \,,
\label{boundbasis2} 
\ee
which also obey the twisted $\widehat{sl(2)}$ algebra 
\be
\left[ Q_n^+ \,, Q_m^- \right] & = &T_{n+m} \,, \non \\
\left[ T_n \,, Q_m^\pm \right] & = & \pm 2 Q_{n+m}^\pm \,.   
\label{bulkalgebraagain}
\ee

We shall see that for the boundary theory (\ref{action}), 
(\ref{boundary}), the integrals of motion are linear combinations of 
(\ref{boundbasis1}) - (\ref{boundbasis2}), and the algebra of these IM 
is a subalgebra of (\ref{bulkalgebraagain}). Before treating the 
general case, it is useful to first consider the special cases of 
``fixed'' ($h \rightarrow \infty$) and ``free'' ($h =0$) boundary conditions.

\vglue 1\baselineskip
\noindent {\bf ``Fixed'' boundary conditions}
\vglue 1\baselineskip

For ``fixed'' boundary conditions ($h \rightarrow \infty$), the 
Fermion fields satisfy
\be
\left( \psi_+ + e^{-i(\phi - \varphi)} \bar\psi_+ \right)
\Big\vert_{x=0} &=& 0 \,, \non \\
\left( \psi_- + e^{i(\phi - \varphi)} \bar\psi_- \right)
\Big\vert_{x=0} &=& 0 \,.
\label{fixed}
\ee
These boundary conditions preserve the $U(1)$ symmetry of the bulk
Lagrangian. 

One can show using the field equations that for ``fixed'' boundary 
conditions there are three linearly independent combinations of the 
charges (\ref{fermichargesbound})  which are integrals of motion: 
\be
\hat Q_{1\ fixed}^{+} &=& 2 \left( 
\bar Q_+ + e^{2i (\phi - \varphi)} Q_+ \right) \,, 
\non \\ 
\hat Q_{1\ fixed}^{-} &=& 2 \left( 
\bar Q_- + e^{-2i (\phi - \varphi)} Q_- \right) \,,
\non \\ 
\hat T_{0\ fixed} &=& 2 T \,.
\label{fixedcharge}
\ee

In analogy with the bulk analysis, we wish to rewrite these IM in 
terms of Fourier modes.  A direct -- but laborious -- approach is to 
substitute into the above expressions the mode expansions 
(\ref{modes}), to use the boundary $S$ matrix to express modes with 
negative rapidity in terms of modes with positive rapidity (see Eq.  
(\ref{boundarySmatrix-definition})), and then to perform the $x$ 
integrals and use certain identities which are obeyed by the boundary 
$S$ matrix.  The result is
\be
\hat Q_{1\ fixed}^\pm & = & 2 \left(
Q_1^\pm + e^{\pm i 2(\phi - \varphi)}Q_{-1}^\pm \right) \,, 
\non  \\
\hat T_{0\ fixed} &= & 2 T_0  \,,
\label{fixedbasisfirst}
\ee 
where $Q_{n}^{\pm}$ and $T_{n}$ are given by Eqs.  
(\ref{boundbasis1}), (\ref{boundbasis2}).  We now observe that the 
same result can be obtained much more readily by assuming that (for 
the purpose of writing a conserved charge in terms of Fourier modes) 
we can make the replacement
\be
\int_{-\infty}^0 dx \rightarrow {1\over 2} \int_{-\infty}^\infty  dx
\label{prescription1}
\ee
in the coordinate-space expression for the charges 
(\ref{fermichargesbound}); and then make in the final result the 
replacement
\be
\int_{-\infty}^\infty d\theta \rightarrow 2 \int_0^\infty  d\theta \,.
\label{prescription2}
\ee
We shall use this device extensively in the work that follows. 

Moreover, we find higher-derivative integrals of 
motion which in terms of Fourier modes are given by
\be
\hat Q_{n\ fixed}^\pm & = & Q_n^\pm + Q_{2-n}^\pm
+ e^{\pm i 2(\phi - \varphi)} \left( Q_{n-2}^\pm 
+ Q_{-n}^\pm \right) \,, 
\qquad n {\hbox{ odd}} \ge 1 \,, \non \\
\hat T_{n \ fixed} &= & T_n + T_{-n} \,, 
\qquad \qquad \qquad \qquad \qquad \qquad \qquad 
n {\hbox{ even}} \ge 0 \,,
\label{fixedbasis} 
\ee
where $Q_{n}^{\pm}$ and $T_{n}$ are given by Eqs.  
(\ref{boundbasis1}), (\ref{boundbasis2}).  (See Appendix A for 
details.)  Notice that the expressions for $\hat Q_{n\ fixed}^\pm$ and 
$\hat T_{n \ fixed}$ are invariant under $n \rightarrow 2-n$ and $n 
\rightarrow -n$, respectively.  The IM for ``fixed'' boundary 
conditions obey the algebra (we now drop the label ``fixed'')
\be
\left[ \hat Q_n^+ \,, \hat Q_m^- \right] & = & \hat T_{n+m} 
+ \hat T_{n+m-4} + \hat T_{2+n-m} + \hat T_{2-n+m} \non  \\ 
& & + 2 \cos(2 (\phi - \varphi)) \left( \hat T_{n-m} 
+ \hat T_{n+m-2} \right) \,, \non  \\ 
\left[ \hat T_n \,, \hat Q_m^\pm \right] & = & \pm  
2 \left( \hat Q_{m+n}^\pm + \hat Q_{m-n}^\pm \right)  \,, 
\label{fixedalgebra}
\ee
and all other commutators vanish.

\vglue 1\baselineskip
\noindent {\bf ``Free'' boundary conditions}
\vglue 1\baselineskip

For ``free''' boundary conditions ($h = 0$), the Fermion fields satisfy
\be
\left( \bar\psi_+ - e^{-i(\phi + \varphi)} \psi_- \right)
\Big\vert_{x=0} &=& 0 \,, \non \\
\left( \bar\psi_- - e^{i(\phi + \varphi)} \psi_+  \right)
\Big\vert_{x=0} &=& 0 \,.
\label{free}
\ee
These boundary conditions break the $U(1)$ symmetry of the bulk
Lagrangian. 

For ``free'' boundary conditions, there are only two linearly 
independent combinations of the charges (\ref{fermichargesbound}) 
which are integrals of motion:
$\hat Q_{1\ free}$ and $\hat Q_{1\ free}^\dagger$, where
\be
\hat Q_{1\ free} = 2 \left( \bar Q_+ + e^{-2i (\phi + \varphi)} Q_- 
- i e^{-i (\phi + \varphi)}T \right) \,.
\label{freecharge}
\ee
Going to Fourier modes, we have
\be
\hat Q_{1\ free} = 2 \left( Q_1^+ + e^{-2 i (\phi + \varphi)} Q_{-1}^- 
- i e^{- i (\phi + \varphi)} T_{0} \right) \,, 
\label{freebasis1} 
\ee 
where $Q_{n}^{\pm}$ and $T_{n}$ are given by Eqs. 
(\ref{boundbasis1}), (\ref{boundbasis2}).

Moreover, we find higher-derivative integrals of motion which in terms 
of Fourier modes are given by
\be
\hat Q_{n \ free} & = & Q_n^+ + Q_{2-n}^+
+ e^{-2 i(\phi + \varphi)}\left( Q_{n-2}^- + Q_{-n}^- \right) \non \\
& & - i e^{- i (\phi + \varphi)}\left( T_{n-1} + T_{1-n} \right) \,,
\qquad\qquad\qquad\qquad\qquad
n {\hbox{ odd}} \ge 1 \,, \non  \\
\hat T_{n \ free} &= & T_n - T_{-n} + T_{-n+4} - T_{n-4} \non \\
& & - 2i  e^{i (\phi + \varphi)}\left( Q_{n-1}^+ + Q_{n-3}^+ 
+ Q_{1-n}^+ + Q_{3-n}^+ \right) \non  \\
& & + 2i  e^{-i (\phi + \varphi)}\left( Q_{n-1}^- + Q_{n-3}^- 
+ Q_{1-n}^- + Q_{3-n}^- \right) \,,
\qquad n {\hbox{ even}} \ge 2 \,, 
\label{freebasis2}  
\ee
together with $\hat Q_{n\ free}^\dagger$. (Here $Q_{n}^{\pm}$ and $T_{n}$ 
are given by Eqs. (\ref{boundbasis1}), (\ref{boundbasis2}). See 
Appendix A for details.) Notice that the expressions 
for $\hat Q_{n\ free}$ and $\hat T_{n\ free}$ are invariant under
$n \rightarrow 2-n$ and $n \rightarrow 4-n$, respectively; and
$\hat T_{n\ free}^\dagger = \hat T_{n\ free}$. The algebra of the IM for ``free'' 
boundary conditions is given by (dropping the label ``free'')
\be
\left[ \hat Q_n \,, \hat Q_m^\dagger \right] & = & \hat T_{n+m} 
+ \hat T_{n-m+2} \,, \non  \\
\left[ \hat T_n \,, \hat Q_m \right] & = & 
2 \Big( \hat Q_{m+n} + 2 \hat Q_{m+n-2} +  \hat Q_{m+n-4} \non \\ 
  & & +  \hat Q_{m-n} + 2 \hat Q_{m-n+2} +  \hat Q_{m-n+4} \Big) 
\,,  \non \\
\left[ \hat T_n \,, \hat Q_m^\dagger \right] & = &  
-2 \Big( \hat Q_{m+n}^\dagger + 2 \hat Q_{m+n-2}^\dagger 
+  \hat Q_{m+n-4}^\dagger \non \\ 
  & & +  \hat Q_{m-n}^\dagger + 2 \hat Q_{m-n+2}^\dagger 
  +  \hat Q_{m-n+4}^\dagger \Big) \,,  
\label{freealgebra}
\ee
and all other commutators vanish.

\vglue 1\baselineskip
\noindent {\bf General boundary conditions}
\vglue 1\baselineskip

A principal result \cite{MN2} is that for general values of 
$h$, the quantities
\be
\hat Q_{1} = 2 \left[ \bar Q_+ + e^{-2i (\phi + \varphi)} Q_- 
- i e^{-i (\phi + \varphi)}\left( 1 - {h^2\over m}
e^{i (\phi - \varphi)} \right) T + {i\over m} e^{-i (\phi + \varphi)}
\psi_-\bar \psi_+\Big\vert_{x=0} \right]
\label{charge}
\ee
and $\hat Q_{1}^\dagger$ are integrals of motion. A proof is 
outlined in Appendix B. As $h$ varies from $0$ to 
$\infty$, the conserved charge $\hat Q_{1}$ interpolates between 
$\hat Q_{1\ free}$ and $\hat T_{0\ fixed}$:
\footnote{In  \cite{MN1}, we exhibited a {\it different} 
set of charges $\hat Q_{1\ old}$, $\hat Q_{1\ old}^\dagger$ which interpolate
instead between $\hat Q_{1\ free}$ and $\hat Q_{1\ fixed}$:
\be
\hat Q_{1\ old} & \buildrel\ h \rightarrow 0 \over \longrightarrow &
 \hat Q_{1\ free} \non \\ 
       & \buildrel\ h \rightarrow \infty \over \longrightarrow &
 -e^{-4i\varphi}\hat Q_{1\ fixed}^\dagger \,. \non 
\ee
The old charges have the drawback of being ``nonlocal'' in time; 
indeed, they are constructed in terms of the quantities $C(\psi_+)$ 
and $C(\bar\psi_+)$, where $C(\psi)$ is defined in Eq. (\ref{C}).} 
%
\be
\hat Q_{1} & \buildrel\ h \rightarrow 0 \over \longrightarrow &
 \hat Q_{1\ free} \non \\ 
       & \buildrel\ h \rightarrow \infty \over \longrightarrow &
  i{h^2\over m}e^{-2i\varphi}\hat T_{0\ fixed} \,. 
\ee
Going to Fourier modes, we have 
\footnote{Note that according to our prescription 
(\ref{prescription1}), (\ref{prescription2}) for going to Fourier 
modes, the boundary term in Eq.  (\ref{charge}) gives no 
contribution.}
\be
\hat Q_1 = \hat Q_{1\ free} + 2 i{h^2\over m} e^{-2i\varphi} T_0 \,,
\label{chargefourier}
\ee
where $\hat Q_{1\ free}$ is given by Eq. (\ref{freebasis1}). 

We also find higher-derivative IM which in terms of Fourier modes are given by
\be
\hat Q_{n} &=& \hat Q_{n\ free} 
+ 4 {h^{2}\over m} e^{-i(3 \varphi + \phi)}
\sum_{k=1}^{n-1\over 2}(-)^{k}
\left( Q_{n-2k}^{-} + Q_{2k-n}^{-}  \right) \non \\ 
& & + 2 i {h^{2}\over m}  e^{- 2 i \varphi}  \left[ 
(-)^{n+1\over 2} T_{2}
+ \sum_{k=1}^{n-3\over 2}(-)^{k+1} \left( 
T_{n-2k+1} + T_{2k+1-n} \right) \right] \non \\ 
& & + 4 {h^{4}\over m^{2}} 
e^{- 4 i \varphi} \left[ 
(-)^{n+1\over 2} \left({n-1\over 4}\right) \hat Q_{1\ fixed}^{-} 
+ \sum_{k=1}^{n-3\over 2}(-)^{k+1} k 
\hat Q_{n-2k\ fixed}^{-} \right] \,,
\qquad n {\hbox{ odd}} \ge 3  \,, \non \\ 
\hat T_{n} &=& \hat T_{n\ free} + 
\left[ 2 {h^{2}\over m}\left(
e^{i(\phi - \varphi)} + e^{-i(\phi - \varphi)} \right) 
- 4 {h^{4}\over m^{2}} \right] \hat T_{n-2\ fixed}  \non \\ 
& & + 4 i {h^{2}\over m}\left[ 
e^{2 i \varphi}\left( Q_{n-1}^{+} + Q_{-n+3}^{+} \right)
- e^{- 2 i \varphi} \left( Q_{n-1}^{-} + Q_{-n+3}^{-} \right) \right] \,,
\qquad n {\hbox{ even}} \ge 2 \,,
\label{generalbasis}
\ee
where $Q_{n}^{\pm}$ and $T_{n}$ are given by Eqs.  
(\ref{boundbasis1}), (\ref{boundbasis2}), 
$Q_{n\ fixed}^{\pm}$ and $T_{n\ fixed}$ are given by Eq.  
(\ref{fixedbasis}), and
$Q_{n\ free}^{\pm}$ and $T_{n\ free}$ are given by Eq.  
(\ref{freebasis2}). (See Appendix B for details.) Note that these IM 
also interpolate between  ``free'' and ``fixed'' IM as $h$ varies 
from $0$ to $\infty$.

These IM obey the following commutation relations:
\be
\left[ \hat Q_1 \,, \hat Q_1^\dagger \right] & = & 2 \hat T_{2} 
+ 4 i {h^{2}\over m}\left( 
- e^{2 i \varphi} \hat Q_{1} + e^{- 2 i \varphi} \hat Q_{1}^{\dagger} 
\right) \,, \non \\ 
\left[ \hat Q_3 \,, \hat Q_1 \right] & = & 
4 i {h^{2}\over m} e^{- 2 i \varphi} \hat Q_{3} 
+ 2 {h^{2}\over m} e^{- i (\phi + 3 \varphi )} \hat T_{2} 
+ 4 i  \left( {h^{2}\over m} e^{- 2 i \varphi} 
- 2 {h^{4}\over m^{2}} e^{- i (\phi +  \varphi )} \right) \hat Q_{1} \,, \non \\ 
\left[ \hat Q_3 \,, \hat Q_1^{\dagger} \right] & = & 2 \hat T_{4} 
- 4 i {h^{2}\over m} e^{2 i \varphi} \hat Q_{3} 
- 2 {h^{2}\over m} e^{i (\phi - \varphi )} \hat T_{2} 
- 4 i {h^{2}\over m} e^{-2 i \varphi} \hat Q_{1}^{\dagger} 
+ 8 i {h^{4}\over m^{2}} e^{i (\phi +  \varphi )} \hat Q_{1} \,, \non \\ 
\left[ \hat T_2 \,, \hat Q_1 \right] & = & 
8 \hat Q_{3} + 8 \left(
1 - {h^{2}\over m} e^{i (-\phi +  \varphi )} \right) \hat Q_{1}
- 8 {h^{2}\over m} e^{- i (\phi + 3 \varphi )} \hat Q_{1}^{\dagger}
\,,  \non \\
\left[ \hat T_2 \,, \hat Q_3 \right] & = &  
4 \hat Q_{5} + 8 \left[
1- {h^{2}\over m} \left( 
e^{i (-\phi +  \varphi )} + e^{- i (-\phi +  \varphi )} \right) 
+ 2 {h^{4}\over m^{2}} \right] \hat Q_{3} \non \\ 
& & + 4 \left(
1 - 2 {h^{2}\over m} e^{i (\phi - \varphi )} \right) \hat Q_{1} 
+ 8 \left( 
{h^{2}\over m} e^{- i (\phi + 3 \varphi )}
- 2 {h^{4}\over m^{2}} e^{- 2 i (\phi + \varphi )} \right) 
\hat Q_{1}^{\dagger} \,, 
\label{generalalgebra} 
\ee
etc. 

\vglue 1\baselineskip
\noindent {\bf Boundary $S$ matrix revisited}
\vglue 1\baselineskip

The IM $\hat Q_{1}$ and $\hat Q_{1}^\dagger$ (as well as the higher 
IM) correctly determine the boundary $S$ matrix.  Indeed, the 
commutation relations of $\hat Q_{1}$ and $\hat Q_{1}^\dagger$ with 
$A(\theta)^\dagger$ are given by (see Eqs. (\ref{chargefourier}),
(\ref{freebasis1}),(\ref{boundbasis1}), (\ref{boundbasis2}))
\be
\left[ \hat Q_{1} \,, A(\theta)^\dagger \right] & = & Z(\theta)\ 
A(\theta)^\dagger  \,, \non \\ 
\left[ \hat Q_{1}^\dagger \,, A(\theta)^\dagger \right] & = & Z'(\theta)\ 
A(\theta)^\dagger  \,,
\ee
where the $2 \times 2$ matrices $Z(\theta)$ and $Z'(\theta)$ are 
given by 
\be
Z(\theta) =2 \left( \begin{array}{cc}
   -c	     & b(\theta)	\\
   a(\theta) & c
	  \end{array} \right)	\,, \qquad \qquad
Z'(\theta) =2 \left( \begin{array}{cc}
   - c^{*}        & a(\theta)	\\
    b(\theta)^{*} & c^{*}
	  \end{array} \right)  \,,
\ee
and
\be
a(\theta) = e^{-\theta} \,, \qquad 
b(\theta) = e^{-2 i (\phi + \varphi) + \theta}  \,, \qquad 
c = i e^{- i (\phi + \varphi)}  - i {h^{2}\over m}  e^{-2 i \varphi} 
\,.
\ee
We next observe that
\be
\hat Q_{1} A(\theta)^\dagger |0\rangle_B 
&=& R(\theta)\ \hat Q_{1} A(-\theta)^\dagger |0\rangle_B 
= R(\theta)\ Z(-\theta)\ A(-\theta)^\dagger |0\rangle_B  
\label{argument1} \\ 
&=& Z(\theta)\  A(\theta)^\dagger |0\rangle_B 
= Z(\theta)\ R(\theta)\ A(-\theta)^\dagger |0\rangle_B  \,.
\label{argument2} 
\ee 
In the first line, we first use the definition (\ref{boundarySmatrix-definition})
of the boundary $S$ matrix, and we then use the commutation relations 
together with the fact $\hat Q_{1} |0 \rangle_{B} = 0$; and in the second line 
we reverse the order of operations. We conclude that
\be
Z(\theta)\ R(\theta) & = & R(\theta)\ Z(-\theta)  \,.
\label{cond1} 
\ee 
An analogous analysis with $\hat Q_{1}^\dagger$ yields
\be 
Z'(\theta)\ R(\theta) & = & R(\theta)\ Z'(-\theta)  \,.
\label{cond2}
\ee
Solving these two relations for $R(\theta)$ leads directly to the 
boundary $S$ matrix given by Eqs. (\ref{boundarySmatrix}), (\ref{iden1}), 
(\ref{iden2}), up to the unitarization factor.

\section{Away from the free Fermion point ($\beta^2 \ne 4\pi$)}

We would like to explicitly construct generalizations of the integrals 
of motion $\hat Q_{1}$ and $\hat Q_{1}^\dagger$ for the 
sine-Gordon model with boundary for $\beta^2 \ne 4\pi$. 

As a preliminary step in this direction, we make the following 
observation. Define the quantities $\hat Q$ and $\hat Q'$
by
\be
\hat Q &=& 2 \left[ \bar\alpha^{-1} \bar Q_+ 
+ \alpha^{-1} {k_+\over k_-}Q_- 
- 2i {e^{-i \xi}\over k_-} \left( {1 - q^{-T}\over q - q^{-1}} 
\right) \right]
\,,  \label{ansatz1} \non \\ 
\hat Q' &=&  2 \left[ \bar\alpha^{-1} \bar Q_- 
+ \alpha^{-1} {k_-\over k_+}Q_+ 
+ 2i {e^{i \xi}\over k_+} \left( {q^T - 1\over q - q^{-1}} \right) \right]
\,, 
\label{ansatz2}
\ee
where $Q_\pm$, $\bar Q_\pm$, and $T$ have commutation relations with 
$A(\theta)^\dagger$ given by Eq. (\ref{comrel}).  A generalization of the 
argument given in Eqs. (\ref{argument1}) - (\ref{cond2}) using $\hat Q$ and 
$\hat Q'$ instead of $\hat Q_{1}$ and $\hat Q_{1}^{\dagger}$, respectively, 
leads to the boundary $S$ matrix $R(\theta)$ given by 
Eq. (\ref{boundarySmatrix}).  For $q \rightarrow 1$, one can verify with 
the help of Eqs.  (\ref{iden1}) and (\ref{iden2}) that the charges 
$\hat Q$ and $\hat Q'$ reduce to the expression (\ref{chargefourier}) 
for $\hat Q_{1}$ in terms of Fourier modes and its Hermitian 
conjugate, respectively.

It remains to construct $\hat Q$ and $\hat Q'$ explicitly in 
terms of the local sine-Gordon field $\Phi(x,t)$, to demonstrate 
their conservation, and to identify their algebra.

\section{Discussion}

For the case $q=1$, we have seen that the ``boundary quantum group'' 
is a one-parameter ($h$) family of infinite-dimensional subalgebras of 
twisted $\widehat{sl(2)}$.  For ``fixed'' ($h \rightarrow \infty$) and 
``free'' ($h =0$) boundary conditions, the subalgebras can be 
described very explicitly.  (See Eqs.  (\ref{fixedalgebra}) and 
(\ref{freealgebra}), respectively.) However, for general values of 
$h$, the subalgebra is evidently more complicated (see Eq.  
(\ref{generalalgebra})), and we have not yet succeeded in giving a 
complete characterization of its structure.  We remark that, to our 
knowledge, the general subject of subalgebras of affine Lie algebras 
remains largely unexplored.

We emphasize that for the general case $q \ne 1$, the most pressing 
problems are to explicitly construct the fractional-spin IM in terms 
of the local sine-Gordon field, and to identify the algebra of these 
IM.  The boundary sine-Gordon field theory (\ref{boundarySG}) can be 
regarded as a free scalar conformal field theory with both bulk and 
boundary integrable perturbations \cite{Z2}, \cite{GZ}.  Within this 
framework, it should be easier to treat the special case of only a 
boundary perturbation (i.e., conformal bulk).  It would also be 
interesting to identify analogues of fractional-spin IM for integrable 
quantum spin chains, which can be solved by the Bethe Ansatz.  A 
related question is whether a systematic construction of the 
fractional-spin IM can be found, in analogy with the so-called quantum 
inverse scattering (QISM) construction of local IM.  Finally, we 
remark that it should be possible to construct fractional-spin IM for 
other integrable field theories with boundaries.

\section{Acknowledgments}

The	work presented here	was	initiated by A.	 B.	 Zamolodchikov,	and
represents an account of a joint ongoing investigation with	him.  We
are	grateful to	him	for	his	invaluable advice, kind	hospitality, and
for	giving us the permission to	present	some of	our	joint results here.
We also thank O.  Alvarez, D.  Bernard, F. Essler, A.  LeClair, 
P. Pearce, V.  Rittenberg, and L.  Vinet for discussions; and we acknowledge 
the hospitality at Bonn University, Rutgers University, and the Benasque 
Center for Physics, where part of this work was performed.  This work 
was supported in part by DFG Ri 317/13-1, and by the National Science 
Foundation under Grant PHY-9507829.

\section{Appendix A: Higher-derivative IM for ``fixed'' and ``free'' 
boundary conditions}

We explain here our strategy for constructing higher-derivative
integrals of motion.  We begin with the case of ``fixed'' boundary 
conditions (\ref{fixed}). The first step is to observe (using the field
equations and boundary conditions) that the following charges are
conserved:
\be
q_{2n+1\ fixed} &=& -{2i\over m^{2n+1}} \int_{-\infty}^0 dx\ \left[
\bar \psi_+^{(n)} {\bar \psi}_+^{(n+1)} +
e^{2i (\phi - \varphi)} \psi_+^{(n)} \psi_+^{(n+1)} \right] \,, \non \\
t_{2n\ fixed} &=& {2\over m^{2n}} \int_{-\infty}^0 dx\ \left[
\psi_+^{(n)} \psi_-^{(n)} + \bar \psi_+^{(n)} {\bar \psi}_-^{(n)} \right] \,,
\label{fixedreducible}
\ee
where $n$ is a nonnegative integer, and ${}^{(n)}$ denotes 
$n^{th}$-order time derivative.  For $n=0$, these IM coincide with 
$\hat Q_{1\ fixed}^+$ and $\hat T_{0\ fixed}$, respectively.  (See Eq.  
(\ref{fixedbasisfirst}).) The second step is to express these IM in 
terms of Fourier modes.  Using the prescription (\ref{prescription1}), 
(\ref{prescription2}), we find that
\be
q_{2n+1\ fixed} &=& 2 \int_0^{\infty} d\theta\ \cosh^{2n}\theta
\left[ e^{-\theta} + e^{2i (\phi - \varphi) + \theta} \right]
A_+(\theta)^\dagger A_-(\theta) \,, \non \\
t_{2n\ fixed} &=&  2 \int_0^{\infty} d\theta\ \cosh^{2n}\theta
\left[ A_+(\theta)^\dagger A_+(\theta)
-  A_-(\theta)^\dagger A_-(\theta) \right] \,.
\ee
The third step is to express these IM in terms of the basis
(\ref{boundbasis1}), (\ref{boundbasis2}) by writing $\cosh \theta$ in
terms of exponentials $e^{\pm \theta}$:
\be
q_{1\ fixed} &=& 2 \left[
Q_1^+ + e^{2 i(\phi - \varphi)}Q_{-1}^+ \right] \,, \non \\
q_{3\ fixed} &=& {1\over 2} \left[ Q_3^+ + Q_{-1}^+ +
e^{2 i(\phi - \varphi)} \left( Q_1^+ + Q_{-3}^+ \right) \right] +
{1\over 2} q_{1\ fixed} \,, \non \\
 &\vdots& \non \\
t_{0\ fixed} &=& 2 T_{0} \,, \non \\
t_{2\ fixed} &=& {1\over 2} \left( T_{2} + T_{-2} \right) 
+ {1\over 2}t_{0\ fixed}  \,,
\non \\
 &\vdots& 
\ee
Finally, it is important to note that the IM $q_{2n+1\ fixed}$ and 
$t_{2n\ fixed}$ for $n > 0$ are not ``irreducible''; i.e., they are 
sums of terms which are separately conserved.  The ``irreducible'' 
conserved quantities evidently are
\be
\hat Q_{n\ fixed}^\pm & = & Q_n^\pm +  Q_{2-n}^\pm +
e^{\pm  2 i(\phi - \varphi)} \left( Q_{n-2}^\pm + Q_{-n}^\pm \right) \,,
\qquad n {\hbox{ odd}} \ge 1 \,, \non \\
\hat T_{n \ fixed} &= & T_n + T_{-n} \,,
\qquad \qquad \qquad \qquad \qquad \qquad \qquad
n {\hbox{ even}} \ge 0 \,,
\ee
which is the result stated in Eq. (\ref{fixedbasis}). 

We follow a similar procedure for constructing higher-derivative IM 
for the case of ``free'' boundary conditions (\ref{free}).  For this case, 
we have instead of (\ref{fixedreducible}) the following ``reducible'' 
conserved charges:
\be
q_{2n+1\ free} &=& -{2i\over m^{2n+1}} \int_{-\infty}^0	dx\	\left\{
\bar \psi_+^{(n)} {\bar	\psi}_+^{(n+1)}
+ e^{-2i (\phi + \varphi)} \psi_-^{(n)}	\psi_-^{(n+1)} \right.	\non \\
& &	\left. \qquad  
- m	e^{-i (\phi	+ \varphi)}	\left[ \psi_-^{(n)}	\psi_+^{(n)}
+ \bar \psi_-^{(n)} \bar \psi_+^{(n)} \right] \right\}	
\,,  \non \\
t_{2n+2\ free} &=& {2\over m^{2n+2}} \int_{-\infty}^0 dx\ \left\{
\bar \psi_+^{(n)'} \bar \psi_-^{(n+1)} + \psi_+^{(n)'} \psi_-^{(n+1)}
+ {m\over 2} \left[ e^{i (\phi	+ \varphi)} \left( 
\bar \psi_+^{(n)} \bar \psi_+^{(n+1)} + \psi_+^{(n)} \psi_+^{(n+1)} \right) 
\right. \right. \non \\
& & \left. \left. \qquad  -  e^{-i (\phi	+ \varphi)} \left( 
\bar \psi_-^{(n)} \bar \psi_-^{(n+1)} + \psi_-^{(n)} \psi_-^{(n+1)} \right) 
\right] \right\} \,,
\label{freereducible}
\ee
where $n$ is a nonnegative integer, and the prime $(\ ' \ )$ denotes 
differentiation with respect to $x$.  Going to Fourier modes and 
expressing the result in terms of the basis (\ref{boundbasis1}), 
(\ref{boundbasis2}), we eventually arrive at the ``irreducible'' IM 
given in Eq.  (\ref{freebasis2}).

\section{Appendix B: IM for general boundary conditions}

In this Appendix we outline a proof that, for general boundary 
conditions, the quantity $\hat Q_{1}$ given by Eq.  (\ref{charge}) is 
an integral of motion.  We also indicate how to construct the 
higher-derivative IM.

We consider first the expression (\ref{charge}) for $\hat Q_{1}$
except {\it without} the boundary term:
\be
P \equiv  \int_{-\infty}^{0} dx\ \Big\{
\bar \psi_+ \dot {\bar\psi}_+  
+ e^{-2i (\phi + \varphi)} \psi_- \dot \psi_-  
+ \left( - m e^{-i (\phi + \varphi)} + h^2 e^{- 2 i \varphi)} \right)
\left( \psi_-\psi_+ + \bar\psi_-\bar\psi_+ \right) \Big\} \,.
\ee
Differentiating with respect to time, we obtain using the field 
equations (\ref{fieldequations}) a sum of boundary terms:
\be
\dot P & = & \int_{-\infty}^{0} dx\  \partial_{1} \left[
\bar \psi_+ {\bar\psi}_+' +  e^{-2i (\phi + \varphi)} \psi_- \psi_-'
+ \left( - m e^{-i (\phi + \varphi)} + h^2 e^{- 2 i \varphi)} \right)
\left( \psi_-\psi_+ - \bar\psi_-\bar\psi_+ \right) \right]  \non \\
& = & \Big\{ 
- \bar \psi_+ \dot {\bar\psi}_+  
+ e^{-2i (\phi + \varphi)} \psi_- \dot \psi_- 
+ h^2 e^{- 2 i \varphi} \left( \psi_-\psi_+ - \bar\psi_-\bar\psi_+ 
\right) \non \\
& & + m \left[ 
\bar \psi_+ \psi_+ + e^{-2i (\phi + \varphi)} \psi_- \bar\psi_-
-  e^{- i (\phi + \varphi)} \left( \psi_-\psi_+ - \bar\psi_-\bar\psi_+ \right)
\right] \Big\}\Big\vert_{x=0} \,,
\label{leftover}
\ee
where the prime $(\ ' \ )$ denotes differentiation with respect to $x$.  
Our objective is to express the result as a time derivative of a local 
boundary term.  We proceed by eliminating $\psi_{\pm}$ in favor of 
$\bar \psi_{\pm}$ using the following identities \cite{MN1} which can 
be derived from the boundary conditions (\ref{BC-interpolating1}), 
(\ref{BC-interpolating2}):
\be
\left[ \psi_+   - e^{-i(\phi + \varphi)} \bar\psi_- 
+ e^{-i \phi} C(\bar\psi_+) \right]\Big\vert_{x=0} & = & 0 \,, \non \\
\left[ \psi_-   - e^{i(\phi + \varphi)} \bar\psi_+
+ e^{i \phi} C(\bar\psi_+) \right]\Big\vert_{x=0} & = & 0 \,, 
\label{identities}
\ee
where the quantity $C(\psi)$ is defined by
\be
C(\psi) = {1\over 1 + {\partial_0\over h^2}}\left( e^{i\varphi} \psi
+ e^{-i\varphi} \psi^\dagger \right) \,.
\label{C}
\ee
We find in this way that
\be
\dot P  =  - \partial_{0} \left( e^{- i \varphi} C(\bar\psi_+) \bar\psi_+ 
\right)\Big\vert_{x=0} 
= \partial_{0} \left( e^{- i (\phi + \varphi)} \psi_- \bar\psi_+ 
\right)\Big\vert_{x=0} \,.
\ee
Therefore, the quantity
\be
\hat Q_{1} \propto P - e^{- i (\phi + \varphi)} \psi_- \bar\psi_+ \Big\vert_{x=0} 
\ee
is an integral of motion.  

In a similar manner, we find the following ``reducible'' higher-derivative IM:
\be
q_{2n+1} &=& -{2i\over m^{2n+1}} \int_{-\infty}^0	dx\	\left\{
\left( \bar\psi_+^{(n)} - h^{2} e^{-2i\varphi} \bar\psi_{-}^{(n-1)} \right)
\left( \bar\psi_+^{(n+1)} - h^{2} e^{-2i\varphi} \bar\psi_{-}^{(n)} \right)
\right. \non \\
& & \qquad + e^{-2i (\phi + \varphi)} 
\left( \psi_-^{(n)}	+ h^{2}  \psi_-^{(n-1)} \right)
\left( \psi_-^{(n+1)}	+ h^{2}  \psi_-^{(n)} \right) \non \\
& & \qquad - m	e^{-i (\phi	+ \varphi)}	\left[ 
\left( \psi_-^{(n)}	+ h^{2} \psi_-^{(n-1)} \right)
\left( \psi_+^{(n)} - h^{2} e^{-2i\varphi} \psi_-^{(n-1)} \right) 
\right.  \non \\
& & \qquad \qquad \qquad \left. \left. + 
\left( \bar \psi_-^{(n)} + h^{2} \bar \psi_-^{(n-1)} \right)
\left( \bar \psi_+^{(n)} - h^{2} e^{-2i\varphi} \bar \psi_-^{(n-1)} \right)
\right] \right\}	
\,, \non \\
t_{2n} &=& -{2\over m^{2n}} \int_{-\infty}^0 dx\ \left\{
\left( \bar \psi_+^{(n)} - h^{2} e^{-2i\varphi} \bar \psi_-^{(n-1)} \right)
\left( \bar \psi_-^{(n)} - h^{2} e^{2i\varphi} \bar \psi_+^{(n-1)} \right)
\right. \non \\ 
& &  + \left( \psi_-^{(n)} + h^{2} \psi_-^{(n-1)} \right)
\left( \psi_+^{(n)} + h^{2} \psi_+^{(n-1)} \right) 
 - {m\over 2} {d\over dt} \left(
\psi_+^{(n-1)} \bar \psi_-^{(n-1)} + \bar \psi_+^{(n-1)} \psi_-^{(n-1)}
\right)  \non \\
& & - {m\over 2} \left[ e^{i (\phi	+ \varphi)} \left( 
\bar \psi_+^{(n-1)} \bar \psi_+^{(n)} + \psi_+^{(n-1)} \psi_+^{(n)} \right) 
-  e^{-i (\phi	+ \varphi)} \left( 
\bar \psi_-^{(n-1)} \bar \psi_-^{(n)} + \psi_-^{(n-1)} \psi_-^{(n)} \right) 
\right] \non \\
& & - m h^{2} \left[ \bar \psi_+^{(n-1)} \psi_-^{(n-1)} 
+ \psi_+^{(n-1)} \bar \psi_-^{(n-1)} 
\right.	\non \\
& &	\left. \left. \qquad + \cos (\phi - \varphi) \left(
\bar \psi_-^{(n-1)} \bar \psi_+^{(n-1)} + \psi_-^{(n-1)} 
\psi_+^{(n-1)} \right) \right]
\right\} \,,
\ee
where $n \ge 1$, and ${}^{(n)}$ denotes $n^{th}$-order time 
derivative.  For $h=0$, these charges reduce to those given in 
Eq. (\ref{freereducible}).  Following the procedure described in 
Appendix A, we obtain the ``irreducible'' IM given in Eq.  
(\ref{generalbasis}).

\vfill\eject

\end{document}